\documentclass[useAMS,usenatbib, usegraphicx]{mn2e}
\usepackage{color}

\title[Dense core compression and fragmentation]
{Dense core compression and fragmentation induced by the scattering
of hydromagnetic waves}
\author[S. Van Loo et al.]
{S. Van Loo$^1$,\thanks{E-mail: svenvl@ast.leeds.ac.uk} 
S.A.E.G. Falle$^2$ and T. W. Hartquist$^1$ \\
$^1$ School of Physics and Astronomy, University of Leeds, Leeds LS2 9JT\\
$^2$ Department of Applied Mathematics, University of Leeds, Leeds LS2 9JT}
\begin{document}

\date{Accepted - . Received - ; in original form -}

\pagerange{\pageref{firstpage}--\pageref{lastpage}} \pubyear{2006}

\maketitle

\label{firstpage}

\begin{abstract}
We have performed 2D hydromagnetic simulations with an adaptive mesh 
refinement code to examine the response of a pre-existing initially
spherical dense core to a non-linear fast-mode wave. One key parameter
is the ratio of the wavelength to the initial core radius. If that 
ratio is large and the wave amplitude is sufficient, significant compression
of the core occurs, as envisaged by \citet{ML98} in their
``turbulent cooling flow" picture. For smaller values of that ratio, 
an initial value of the ratio of the thermal pressure to magnetic pressure
of 0.2, and sufficiently large wave amplitude, the scattering induces
the production of dense substructure in the core. This substructure
may be related to that detected in the dense core associated with 
the cyanopolyyne peak in TMC-1. 
Our simulations also show that short-wavelength waves, contrary to 
large-wavelength waves, do not confine dense cores.
\end{abstract}


\section{Introduction}
\citet{BM02}, \citet{PN02}, \citet{Getal03}, \citet{Letal04}, \citet{NL05},
and \citet{V05} are amongst the authors who have performed 3D simulations
to investigate the formation of clumps in media in which the value of $\beta$,
the ratio of the thermal pressure to magnetic pressure, is initially small
everywhere but in which velocity perturbations exist. These calculations
have included many ingredients.

In three previous papers, we have carried out 1D (\citealt{FH02}, hereafter
FH02; \citealt*{LFH05}, hereafter LFH05) and 2D (\citealt{VFH06}, hereafter
VFH06) numerical studies of MHD waves in low-$\beta$ plasma to elucidate
the mechanism by which clumps form. We consider the results of particular
relevance to the formation of dense cores (e.g. \citealt{Cetal02}) within
translucent clumps like those discovered in CO maps of the Rosette Molecular
Cloud \citep{WBS95}. The FH02 and VFH06 studies concerned small-amplitude
waves and involved 1D and 2D calculations, respectively. A significant
difference in the results of the two studies is that dense core-like 
structures persisted longer in the 2D simulations than in the 1D simulations.
In 2D, such a structure survived as long as the amplitude that might 
be loosely associated with the fast-mode component of the large-scale 
perturbation remained a significant fraction of its initial value.

The interaction  of   fast-mode  waves   with   large-scale  density
inhomogeneities produced sub-structure  within the large-scale clumps.
The  large-scale structures  themselves  arise due  to the  non-linear
steepening of the fast-mode waves. The present paper is concerned with
a  related   but  somewhat  different  problem:  the   response  of  a
pre-existing  dense core-like  structure  to a  background wave.   
The effect of large-amplitude hydromagnetic waves has already 
been studied analytically by \citet{MZ92}. Using a time-averaged form 
of the virial theorem, they showed that a turbulent medium can 
pressure-confine a dense clump. A limitation of their analysis 
is that it does not describe the effect of large-wavelength 
perturbations. With our numerical model, we can study the effect of 
both large and small wavelength perturbations.

Numerical simulations of this problem are limited, as most studies 
(e.g. \citealt{MMK94}) only describe the effect of a shock hitting 
an inhomogeneity. The flow initially compresses the dense core, but eventually 
destroys and fragments the core. A similar behaviour can also be 
expected for non-linear waves. Our simulations can be considered 
a natural extension and generalisation of work on shocks interacting with 
inhomogeneities.

The response of a dense core to a background wave may well be 
relevant to the detection of substructure in Core D,
which coincides with the cyanopolyyne peak, in the TMC-1 by \citet{P98}.
TMC-1 is a ridge of dense cores, each of which has a size of 0.1 -- 0.2 pc
and a mass of one to several solar masses (e.g. \citealt{H92}). The 
substructures studied by \citet{P98} have sizes of 0.03 - 0.06 pc and 
masses of 0.01 - 0.15 $M_{\sun}$. Most are not bound by self-gravity
and are too small to produce brown dwarfs. Since many of these substructures
have masses below the Jeans mass, they must have formed by some
mechanism other than gravitational instability.

A possible mechanism for forming the substructure may be the same as or
related to the mechanism giving rise to dense core formation in the
picture that we have pursued in FH02, LFH05 and VFH06 and which is, 
no doubt, contributing to dense core formation in the other numerical 
simulations referenced above. However, that mechanism ceases to give large 
density contrast structures if $\beta \geq 1$ unless the amplitude of the 
wave is hyper-Alfv\'enic (FH02, LFH02). Some dense cores may have values of
$\beta$ as low as 0.1 (e.g. \citealt{WT00}, \citealt{WT02})  but some 
may have values of $\beta$ greater than unity (e.g. \citealt{KWC06}).
The lines emitted by species such as CS, NH$_3$ and CO in dense cores show
substantial non-thermal components to their broadening, which are, however,
usually not highly supersonic with respect to the thermal sound speed in
H$_2$ (e.g. \citealt{FM92}). For the substructure in Core D of TMC-1, the 
non-thermal line broadening component is important but in most substructures
markedly subsonic in the above sense. The fact that dense cores possess 
substantial non-thermal internal velocities and, in some cases, have values
of $\beta$ somewhat below unity supports the possibility that wave scattering
on a dense core may lead to substructure like that of TMC-1's Core D.

In the present paper we 
examine the effect of MHD waves on dense cores. Additionally, we derive
the properties of the waves that produce substructures within dense cores.
The paper is structured as follows. In Sect.~\ref{sect:model}
we describe the numerical model. We then describe how the dense core 
evolves in the absence of MHD waves (Sect.~\ref{sect:noMHDwave})
in order to understand the effect of the waves (Sect.~\ref{sect:mhd wave}). 
Finally, we summarise and discuss these results in 
Sect.~\ref{sect:conclusions}.

\section{Numerical model}\label{sect:model}
We can use different approaches to study the effect of MHD waves on a
dense core. One way is to continuously excite a wave at the boundary of the 
computational domain. The wave then propagates to the core and interacts 
with it. Another option is to embed a dense core in a wave. In our 
calculations we use the latter as dense cores arise where the waves are.

An initially uniform quiescent medium is perturbed by a magnetosonic wave.
We assume that the background magnetic field is in the $x-y$ plane and that 
wave is propagating in the positive $x$-direction. Furthermore, 
we work in units so that the unperturbed background state is given by 
\[
\rho = 1, {\quad} {\bf v}=0, {\quad} B_{x}=1,
{\quad} B_{y} = \alpha.
\]
The gas pressure $p_g$ is given by an isothermal equation of state, i.e.
$p_g = \rho a^2$ where $a$ is the isothermal sound speed. In our 
calculations, we use $a = 0.1$ unless otherwise stated.

We superpose a fast-mode magnetosonic wave on the background state. 
We do not examine slow-mode waves, as they only generate high-density
contrasts if their initial density perturbation is already large. 
Contrary to FH02 and VFH06, 
we do not only consider small-amplitude waves but also finite-amplitude waves.
Then it is no longer possible to calculate the initial state using 
the linear approximation of the wave. Therefore, we use the approach of 
LFH05. For a fast-mode wave, 
the form of the wave written in terms of the primitive variables
\[
{\bf P} = [\rho, v_x, v_y, B_x, B_y]^t,
\]
must satisfy 
\[
{{\partial {\bf P}} \over {\partial  x}} \propto {\bf r}_f \equiv 
	\left\{\begin{array}
		{l@{\quad}l}
		\left(\rho, c_f, -\frac{c_f B_x B_y}{\Delta_f}, 0, 
			\frac{\rho c_f^2 B_y}{\Delta_f}\right) & B_y \neq 0\\
		\left(0, 0, 1, 0, -\sqrt{\rho}\right)& B_y = 0
		\end{array} \right.
\]
where $r_f$ is the right eigenvector for a fast wave propagating in the 
positive $x$-direction, $c_f$ the fast magnetosonic speed and 
$\Delta_f = \rho c_f^2 - B_x^2$. Given the $x$-dependence of one
of the primitive variables, the others can be determined from the above 
expression.

As the $x$-component of the velocity disappears when $B_y =0$, it is 
convenient to specify the profile of $v_y(x)$. We assume that the 
$y$-component of the velocity associated with the wave varies sinusoidally 
with $x$ and with wavelength $\lambda$, i.e. 
\[
        v_y = A_y \sin\left(\frac{2\pi (x - x_l)}{\lambda}\right),
\]
where $A_y$ is a constant and $x_l$ the left boundary of the 
computational domain. $A_y$ is chosen so that the maximum 
value for the total velocity ($\sqrt{v_x^2+v_y^2}$) is equal to a given  
amplitude $A$. The initial state at a position  $x +\Delta x$ and $y$ 
can then be calculated using 
\[
        {\bf P}(x+\Delta x, y, 0) = {\bf P}(x, y, 0) -
        \left(\frac{{\bf r}_f \Delta_f}{c_f B_x B_y}
        \frac{\partial v_y}{\partial x}\right)(x) \Delta x,
\]
where the right hand side is evaluated at $x$.

We still need to embed a dense core in 
the fast-mode wave. We represent the core by a uniformly dense 
region of radius $r$ at the centre of the computational box.  
The density $\rho_0$ of the core is taken such that $\beta$  
within the core is between 0.1 and 1 as polarisation observations
suggest (\citealt{WT00}; \citealt{WT02}) for some dense cores. 
As $\beta = 2\rho a^2/B^2$, 
the core density must lie between 5 and 50 for $a = 0.1$. Note that 
there is a small variation of $\beta$ within a dense core as the magnetic 
pressure changes across the core. 

Our model can then be  specified by the isothermal 
sound speed $a$, the value $\alpha$ for background $B_y$ field,  the 
wave properties (i.e. the wavelength $\lambda$ and the amplitude $A$),
and the core properties (i.e. the density $\rho_0$ and radius $r$).
However, as we use dimensionless units, the radius of the core 
and the wavelength of the fast-mode wave are not independent parameters. 
The relevant free parameter is $\lambda/r$. Therefore, we assume a fixed 
core radius of $r =2$ in all our calculations and only vary $\lambda$. 

To solve the 2D MHD equations we use an adaptive 
mesh refinement code. The basic algorithm is a second order 
Godunov scheme which uses a linear Riemann solver \citep{F91}.
We included the divergence cleaning algorithm of \citet{D02}
to stabilise the numerical scheme.
The code uses a hierarchy of grids such that the grid spacing of 
level $n$ is $\Delta x/2^n$ where $\Delta x$ the grid spacing of the 
coarsest level. A refinement criterion determines where in 
the computational domain a finer grid is needed. The computational 
domain is $-10 < x < -10$ and $-10 < y < 10$ and we use periodic
boundary conditions. The finest grid has a resolution of 800$\times$800.

\section{Evolution of a dense core in the absence of MHD waves}
\label{sect:noMHDwave}

In an isothermal gas a dense core is overpressured, which  means that the 
core expands. As this introduces waves and shocks into the background 
medium even when the core is not embedded in a wave, it is convenient to 
discuss this expansion in the absence of an embedding wave. This problem 
is the magnetic version of the 2D isothermal Riemann problem. 

\begin{figure}
\includegraphics[width = 8.4 cm]{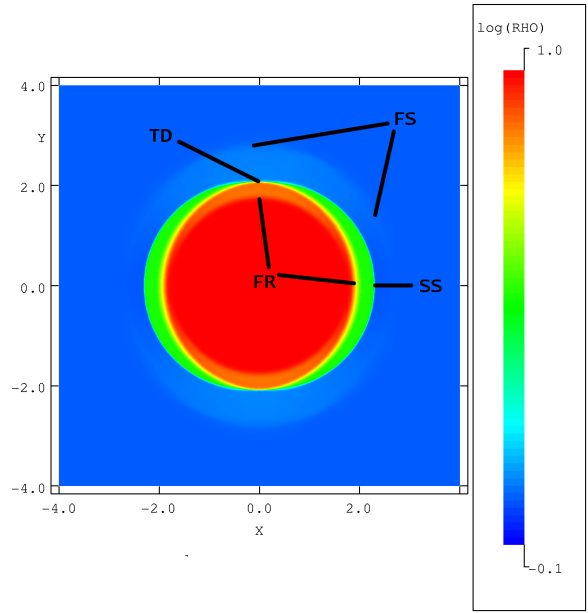}
\caption{The density structure of an expanding dense core. The expansion 
	depends strongly on the angle between the propagation direction and 
	the magnetic field. Perpendicular to the field a fast-mode shock (FS)
	propagates outward and a fast-mode rarefaction (FR) inward. These 
	waves are separated by a tangential discontinuity (TD). For expansion 
	at oblique angles the TD breaks up into a slow-mode shock (SS) and 
	a slow-mode rarefaction (SR). Parallel to the field the FS 
	disappears. The model parameters are
	$\rho_0 = 10$, $a = 0.316$ and $\alpha = 0$. The core is not embedded 
	in a wave.}
\label{fig:core_without_wave}
\end{figure}

In a purely hydrodynamical case, a shock develops and propagates outward 
into the rarefied gas. At the same time, a rarefaction wave reduces the 
density inside the core by travelling inward. As there is no preferred 
direction, the expansion is isotropic. However, this is no longer true in
a magnetised medium because waves (and, thus, also shocks) have a propagation 
speed which depends on the angle between the magnetic field and the direction 
of propagation. Figure~\ref{fig:core_without_wave} 
shows the expansion of a dense core in a magnetic field. 

Slow-mode waves or Alfv\'en waves cannot propagate
perpendicular to the magnetic field. Thus, only fast-mode features develop
in those directions.  A weak fast-mode shock is propagating outward into 
the background medium, while a 
fast-mode rarefaction wave propagates into the dense core. The magnetic 
pressure increases in material through which the shock and rarefaction 
wave have passed.  A tangential discontinuity\footnote{A discontinuity is 
a tangential discontinuity if there is neither magnetic 
flux nor mass flux across it, i.e. $u_n = 0$ and $B_n = 0$ \citep{Burgess}.}
exists between the material that has passed through the front of 
the rarefaction wave and  the material that has passed through the shock.
The total pressure ($p_g + B^2/2$) is continuous at the tangential 
discontinuity but neither $p_g$ or $B^2/2$ is. The fast-mode shock is 
sufficiently weak that 
it does not produce a large density discontinuity. Hence, the largest jump
in density is at the tangential discontinuity. As it coincides with the edge of 
the expanding dense core, the tangential discontinuity moves outward with the 
gas swept up by the fast-mode shock. 

%Expansion 
%perpendicular to the magnetic field is slow relative to {\tt Add or change}.

Slow-mode waves do exist for directions oblique and parallel to the magnetic 
field. A slow-mode shock moves outward and a slow-mode rarefaction wave 
propagates inward. 
These slow-mode features lie between the fast-mode shock and rarefaction wave. 
Although the tangential discontinuity disappears, there is still a sheet 
in between the slow-mode shock and rarefaction wave separating the initial 
core material and the shocked background material.

The density of the material within a region bounded by the 
slow-mode shock is comparable to that of the 
dense core material that has passed through the slow-mode rarefaction wave,
which would make the observational identification of the dense-core boundary
difficult. Since the slow magnetosonic 
speed increases when the angle with the magnetic field decreases, the 
expansion of the core is faster along the magnetic field lines than
perpendicular to it. 

In our calculation the fast-mode shock weakens when the 
angle between its direction of propagation and the magnetic field 
decreases. For parallel expansion the fast-mode shock even turns 
into a linearised wave. Also, the slow-mode rarefaction wave merges with the 
fast-mode rarefaction (see Fig.~\ref{fig:core_without_wave}). 

Finally, the core disperses as there is no force holding it 
together. The dispersal is faster for higher initial 
densities of the core. Also, when the thermal gas pressure 
becomes more dominant in the background gas (i.e. $\beta$ increases),
the dense core disappears more quickly.

\section{A dense core embedded in an MHD wave}\label{sect:mhd wave}

When a dense core is embedded in a fast-mode wave, density substructures 
can be generated by the interaction between the wave and the core. The analysis
in FH02 shows that high-density contrasts are associated with slow-mode 
waves when $\beta$ is small. They also showed that non-linear steepening of a 
fast-mode wave can readily excite these slow-mode components. 

To produce substructures, slow-mode waves must be excited within the core 
itself (as their propagation speed is low). This means that the 
time-scale for non-linear 
steepening $t_{ns} \approx \lambda/2 A$ must be smaller or comparable
to the Alfv\'en crossing time $t_A = r/c_A$. We immediately see that 
for large wavelengths $t_{ns} \gg t_{A}$ even if their amplitude is 
comparable to the Alfv\'en speed. Short-wavelength waves, however,
steepen within the core itself as long as their amplitude is not too
small. 

The interplay between MHD waves and a dense core is, thus, wavelength-dependent.
Therefore, it is convenient to discuss the results for long and short 
waves separately. We typically take $\lambda = 20$ and $\lambda =2$ 
in our calculations.

\subsection{Results for large wavelengths} 
\begin{figure}
\includegraphics[width = 8.4 cm]{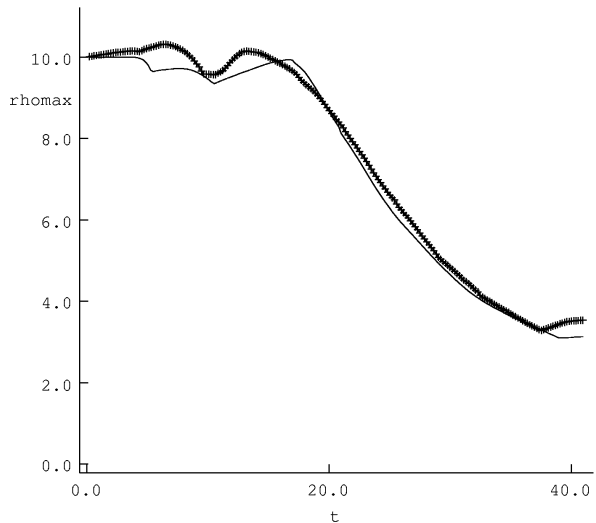}
\caption{The temporal evolution of the maximum density within the dense core.
	The thin solid line shows the evolution in a homogeneous background 
	gas (i.e. $A = 0$), while the thick solid line represents a model 
	with a small-amplitude wave.
	The model parameters are: $\rho_0 = 10$, $\alpha =0.25$, 
	$\lambda = 20$, $A = 0.05$ and $a = 0.1$.  Initially, $\beta = 0.2$ 
	in the dense core.}
\label{fig:l20_sa}   
\end{figure}
Although the time-scale for non-linear steepening for large-wavelength
fast-mode waves is too long for exciting slow-mode components within the dense 
core, the wave strongly affects the evolution of the core.
The most direct effect of the core on a fast-mode wave is the reduction of 
the propagation speed within the core because $c_f \approx  B/\sqrt{\rho}$ 
(for $a < c_a$). This means that a part of the wave - i.e. the part 
within the core - moves slower than the rest of the wavefront. Consequently,
the velocity of the gas must change at the boundary of the core.

However, more importantly, this induces a change in the magnetic field 
at the boundary  of the core, which follows directly from the induction 
equation 
\[
	\frac{\partial {\bf B}}{\partial t} = \nabla \times ({\bf v} 
		\times {\bf B}).
\]
More specifically, $B_x$ increases at the top region of the core and 
decreases at the bottom (or inversely depending on the sign of the
velocity changes).  The $y$-component of the field, $B_y$, shows a similar 
behaviour, but here left and right of the core. 

\begin{figure}
\includegraphics[width = 8.4 cm]{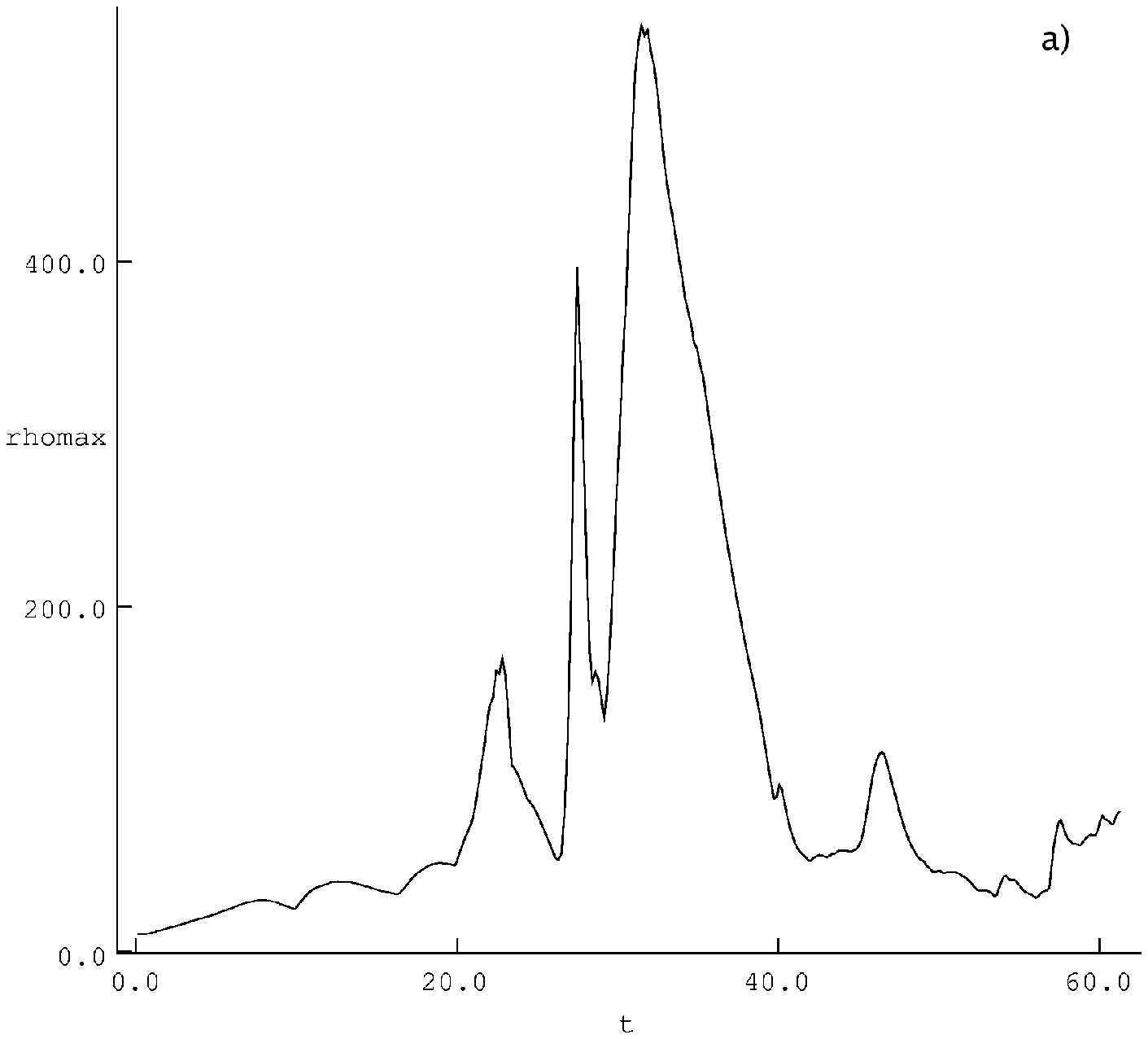}
\includegraphics[width = 8.4 cm]{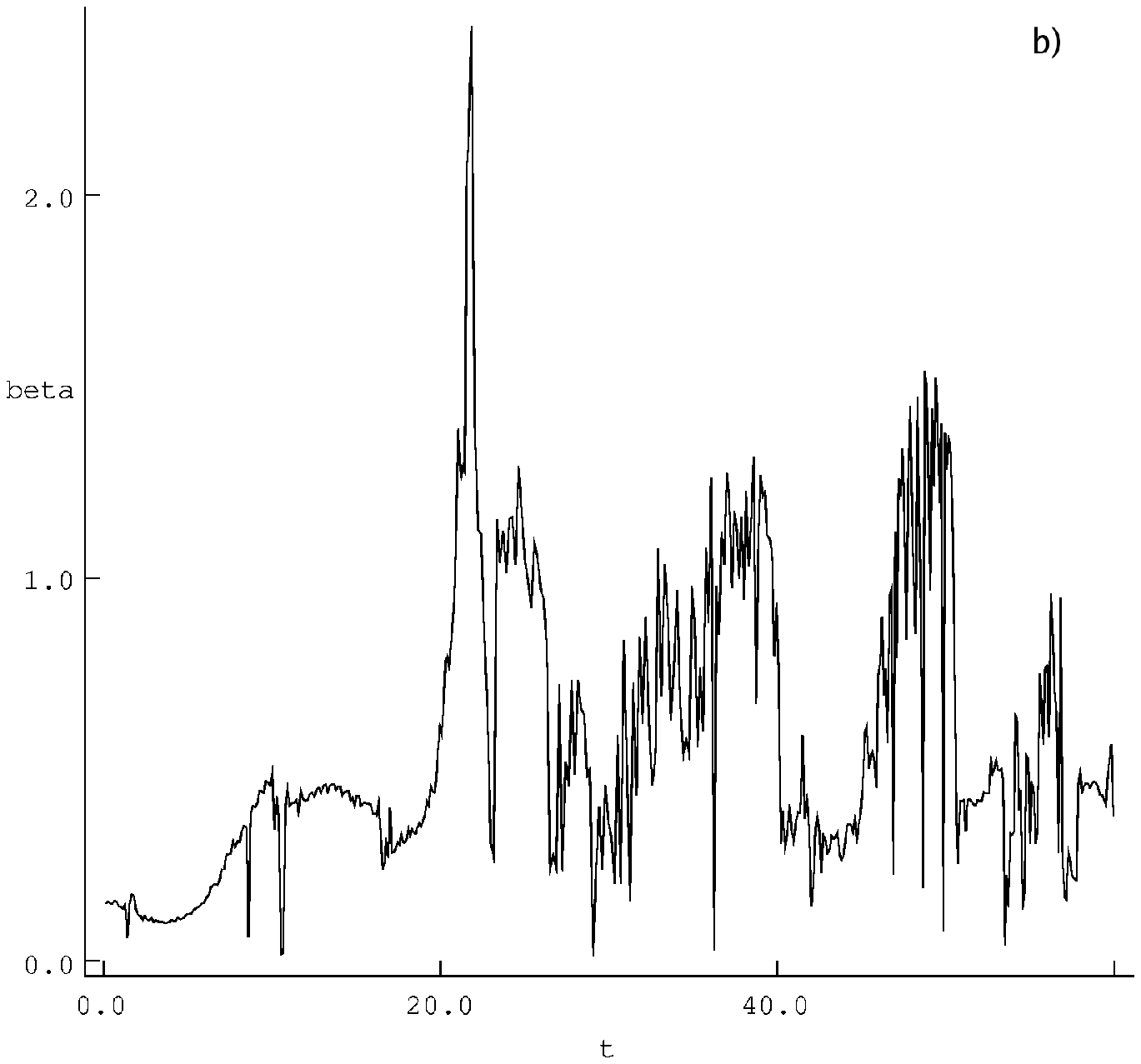}
\caption{(a) Similar to Fig.~\ref{fig:l20_sa}, but now for 
	a large-amplitude fast wave. (b) The plasma beta associated with 
	the fluid element that has the highest density. 
	The model parameters are given by $\rho_0 = 10$, $\alpha = 0.25$, 
	$\lambda = 20 $, $A =0.5$ and $a = 0.1$.}
\label{fig:max_dens1}
\end{figure}

Thus, the interaction with the wave causes an increase of the magnetic 
pressure. While there is initially no discontinuity in the total pressure 
perpendicular to the magnetic field (see Sect.~\ref{sect:noMHDwave}), 
the total pressure outside the core can be much higher 
than inside the core. The magnitude of this change depends on the amplitude 
of the fast-mode  wave. Small-amplitude waves are too weak to produce any 
significant change and the core then disperses as though no fast-mode wave 
is present in the background gas (see Fig.~\ref{fig:l20_sa}).
However, when the amplitude of the fast-mode wave is large, the magnetic 
pressure 
increases significantly. Now the region outside the core is overpressured
compared to the core itself.
A fast-mode shock arises and propagates into the core. 
This shock is weak and, therefore, compresses the gas by a small 
factor as can be seen in Fig.~\ref{fig:max_dens1}a.

Although the expected increase in density from the motion of the wave
across the core is small, Fig.~\ref{fig:max_dens1}a shows that the 
maximum density increases significantly for $t > 20$. We also find that most 
mass is concentrated within a single region. As the time-scale corresponds
to the time for a wave to steepen into a shock ($t_{ns} \approx 20$
for $\lambda = 20$ and $A = 0.5$), we expect that this high-density structure 
is generated by the interaction between the fast-mode shock and the core 
exciting slow-mode waves. However, $\beta$ needs to smaller than unity for 
this process to be effective (see the analysis in FH02). 
Figure~\ref{fig:max_dens1}b shows that $\beta$ for the fluid element 
associated with the maximum density obeys this constraint reasonably well.

We can now examine how the model parameters influence the wave-core interaction.
Of all parameters the wave amplitude 
is perhaps the most important. The wave amplitude determines how 
much the magnetic pressure changes at the boundary. Small-amplitude
waves do not affect the evolution of the core, while large-amplitude waves
modify the entire core structure. In the latter case, a core can be 
compressed to a density considerably higher than the initial one.
The maximum density is limited by the $\beta \la 1$ constraint.
Because of this constraint, it is also more difficult to produce 
high-density contrasts in models where $\rho_0$ and $a$ are high. 
Then $\beta$ is already close to unity or higher.

When the wavelength is larger than the radius of the core, the initial 
position of the core centre with respect to a wave node
becomes an additional free parameter. However, the fast-mode
wave still propagates with a lower speed inside the core. The variations
described above always occur. Hence, the results are essentially independent 
of the initial position.

\subsection{Results for small wavelengths} 
A small-wavelength wave steepens much faster into a shock than one with 
a large wavelength.
If the amplitude of the small-wavelength wave is not too small, the time-scale 
for non-linear steepening is shorter than the Alfv\'en crossing time.
This means that a shock arises before the wave leaves the dense core. 
Then the generation of density substructures can be readily described 
with the models of FH02 and VFH06.

When a fast-mode wave steepens into a shock, slow-mode waves are excited. This 
produces persistent perturbations with large density contrasts. The fast-mode
wave is most efficient in exciting the slow-mode components for modest values
of $\alpha$, which is the tangent of the angle between its direction of 
propagation and the magnetic field. Steepening is slower for very small 
values of $\alpha$ and the shock does not produce slow-mode waves, 
when $\alpha$ is too large (FH02). 

\begin{figure}
\includegraphics[width = 8.4 cm]{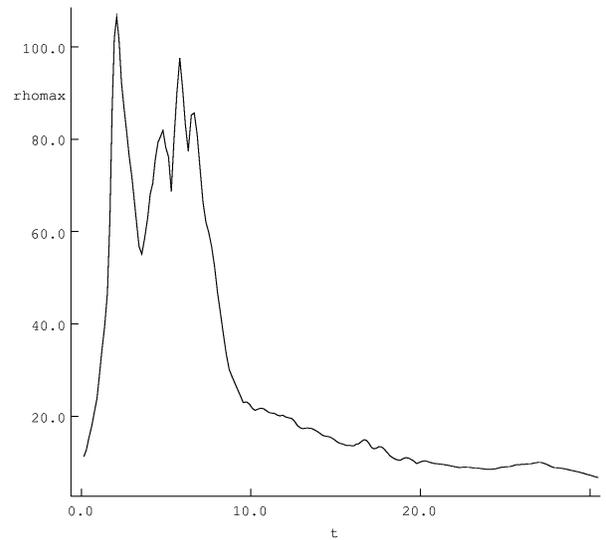}
\caption{Similar to fig.~\ref{fig:max_dens1}, but now for $\lambda = 2$. }
\label{fig:max_dens2}
\end{figure}

\begin{figure}
\includegraphics[width = 8.4 cm]{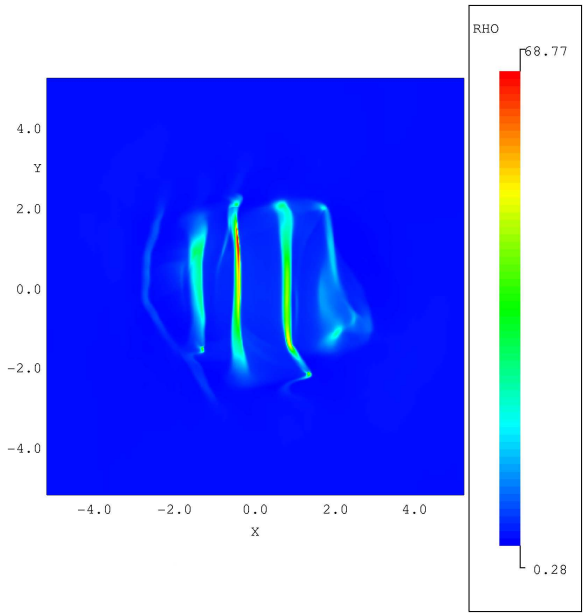}
\caption{The density structure at $t =3$ inside the dense core.
The high-density regions are produced due to non-linear steepening  
of the fast-mode wave. The initial state is given by 
$\lambda =2$, $A = 0.5$, $T_e = 0.01$, $\alpha = 0.25$ and $\rho_0 = 10$.}
\label{fig:la_l2}
\end{figure}

The maximum density due to the non-linear steepening is roughly 
\[
	\rho_{max} = \frac{A^2 \rho_0}{a^2}.      
\]
Small-amplitude waves do not change the density in the core significantly. 
The dense core then disperses as in the case of small-amplitude 
large-wavelength waves,
i.e. as if the core is not embedded in a wave. Large-amplitude waves, however,
generate high-density regions within the core (see Fig.~\ref{fig:max_dens2}). 
These density perturbations are aligned with the $y$-axis as can be seen in 
Fig.~\ref{fig:la_l2}. This is because, except close to the boundary,
there is little variation in the $y$-direction inside the core.

\begin{figure}
\includegraphics[width = 8.4 cm]{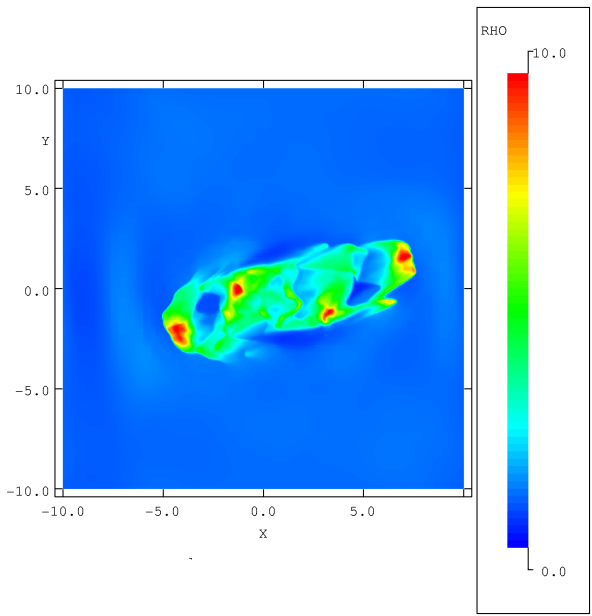}
\caption{Similar to Fig.~\ref{fig:la_l2}, but now for $t =18$.
Dense substructures form within the expanding core due to the interaction
between dense regions and the initial fast mode.}
\label{fig:l2_t15}
\end{figure}

The density perturbations generated by non-linear steepening subsequently
decay. This happens on a time-scale (see FH02) 
\[
	t_e \approx \frac{\lambda}{A}.
\]
Although the dense regions expand and disperse, high-density regions can 
arise for some time due to the interaction of the dense regions with the 
initial fast-mode wave (see VFH06). This interaction excites slow-mode
waves and produces dense substructures. Since the sheet-like structures 
disperse, the core has a more homogeneous density structure with small-scale 
structures nested inside. This can be seen in Fig.~\ref{fig:l2_t15}. 
Also note that, although large density contrasts are generated within the 
core, the core itself still exists as an entity. 
However, because of the fragmentation of the core and associated
motions, the kinetic energy inside the core contributes significantly 
to the total energy. The dense core, therefore, expands more rapidly than
when there is no scattering wave.

High-density structures do not only arise inside the core, but also at 
the boundary. In the explanation above, we neglected the effects occurring
at the boundary of the core. In a similar way as for the large-wavelength 
waves, 
overpressured regions arise producing fast-mode shocks which move into 
the core. Because of the compression caused by the fast-mode shock, 
dense substructures, thus, not only form inside the core but also at 
the boundary of the core. 

\section{Summary and discussions}\label{sect:conclusions}
We have studied the interaction between magnetosonic waves and dense cores
for which the ratio, $\beta$, of thermal gas pressure to magnetic pressure 
is initially 0.2. We follow the evolution of a core embedded in a 
fast-mode wave. We find that large-amplitude waves can change the 
evolution of the core considerably. 

The interaction between the fast-mode wave and a dense core depends strongly
on the relative size of the core to 
the wavelength. When the core radius is smaller than the wavelength, 
the strongest effects are induced near the boundary of the core. Eventually, 
these effects disrupt the global structure of the core, but confines it. 
Furthermore, we find that a core can be compressed significantly. This result 
suggests that a large-wavelength fast-mode wave can trigger the collapse of a 
gravitationally unbound dense core, which is of relevance to the turbulent
cooling flow picture of \citet{ML98}. However, additional calculations 
which include self-gravity are required to confirm this. 

Short-wavelength waves, on the other hand, play an important r\^ole in the 
generation of substructure in a core without breaking up the core. 
However, contrary to the large-wavelength waves, the dense core expands 
faster than without the scattering wave. The wave thus does not confine 
the core.
Although $\beta$ is close to unity in our simulations, slow-mode waves 
excited by the non-linear steepening of the fast-mode wave produce high-density 
contrasts within the core. Such small-scale features have been observed in 
the cores of cold dense clouds such as TMC-1 \citep{P98}. We do not 
find the large number of microclumps inferred from observations of Core D, but
this can be readily explained by the limitations of our model.
The most important limitation is perhaps that we only studied the effect 
of a single wave. A real velocity field is more accurately 
described as a superposition of an ensemble of waves. Due to wave-wave 
interactions, large structures then break up into smaller ones.

\citet{HWV01} showed that the substructures in Core D need to be about
0.1 Myr old to comply with the observed molecular variety and abundances.  
If the substructures are being generated due to slow-mode excitation,
the timescales should be of the same order. For a core radius of $R =0.1$~pc 
and $a = 0.3~{\rm km~s^{-1}}$ and a somewhat larger Alfv\'en speed, 
the substructures in our
simulations arise on timescales of a few $10^5$ yr (a few times
$R/c_a$ which is the relevant time-scale in our simulations).
As both timescales are of the same order, the microstructure in dense 
cold core can indeed be generated by small-wavelength magnetosonic waves.

\section*{Acknowledgements}
SVL gratefully acknowledges the financial support of PPARC.

\label{lastpage}

\end{document}